\documentclass[a4paper,10pt]{article}

\usepackage[left=2.8cm,right=2.8cm,top=2.6cm,bottom=2.6cm]{geometry}

\usepackage{amsfonts,amssymb,amsthm,amsmath}
\usepackage{graphicx}
\usepackage{dcolumn}
\usepackage{bm}
\usepackage{xcolor} 

\renewcommand\d{\mathrm{d}}
\newcommand\Round[1]{{\left(#1\right)}}
\newcommand\Square[1]{{\left[#1\right]}}
\newcommand\of[1]{\Round{\textstyle#1}}

\title{A Drift-Diffusion Approach for Pearson Type IV Heavy-Tail Distributions}

\author{Daniele Belardinelli\thanks{daniele.belardinelli@inaf.it} \and Simone Benella \and Mirko Stumpo \and Giuseppe Consolini\thanks{giuseppe.consolini@inaf.it}}%

\date{}

\begin{document}

\maketitle

\begin{abstract}
Non-Gaussian stationary fluctuations with heavy tails and finite-order moment cutoffs are ubiquitous in non-equilibrium systems. Here, we derive the Pearson type IV distribution as the stationary solution of a standardized Fokker-Planck equation with linear drift and quadratic diffusion. We establish necessary and sufficient conditions for its emergence and show that the linear-quadratic structure, although sufficient, is not necessary. We also give explicit closed expressions for the cumulative distribution function, a moment recurrence relation, skewness, kurtosis, and mode. From a physical perspective, our results reveal that heavy tails and asymmetry originate from distinct features of state-dependent diffusion: its local nonlinear curvature and gradient, respectively. In particular, asymmetric stationary fluctuations can emerge even in the presence of a symmetric linear restoring drift. This framework provides a general stochastic description of asymmetric heavy-tailed fluctuations in non-equilibrium systems, including space plasma turbulence.
\end{abstract}

\section{\label{sec:level1}Introduction}

The statistical characterization of fluctuating observables in non-equilibrium systems is a central and long-standing problem at the intersection of stochastic processes, kinetic theory, and statistical mechanics \cite{bertini2015macroscopic, touchette2009large}. 

In many physical contexts (from the velocity and vorticity fields of turbulent flows to the fluctuating forces in driven colloidal suspensions, the displacement statistics of particles undergoing anomalous diffusion, the current fluctuations in driven lattice gases and active matter systems, and even financial time series) the empirically observed probability distributions depart significantly from Gaussianity, exhibiting a leptokurtic shape (heavy tails), finite-order moment cutoffs, and pronounced asymmetry \cite{chechkin2017brownian, metzler2020superstatistics, akimoto2026anomalous, cont2001empirical, li2005origin}. Furthermore, this non-Gaussian statistics is not a consequence of strongly driven or chaotic regimes, but it emerges in transport processes characterized by long-range correlations or memory effects, in systems near critical points where fluctuations become scale-free, and in heterogeneous media where spatial disorder couples to the dynamical degrees of freedom. 

A possible framework for characterizing the tails of such distributions is provided by the Large Deviation Theory (LDT), which goes beyond the central limit theorem by quantifying the probability of rare but significant fluctuations \cite{touchette2009large}. The LDT rate function, also named \textit{Cramer's function}, is the quantity that contains the information on the decay of probabilities for large events. Indeed, if the rate function is not quadratic as it occurs for equilibrium systems when central limit theorem (CLT) is valid, it can generate heavier-than-Gaussian tails. Furthermore, rate function is deeply connected to thermodynamic potentials and entropy production in non-equilibrium systems.

Another possible framework for the emergence of leptokurtic probability distribution functions is the non-extensive statistical mechanics \cite{abe2001nonextensive,umarov2022mathematical}. In this case the assumption of a loss of the extensive character of the entropy, due for instance to the emergence of long-range correlations, can generate fluctuations whose statistics departs from the CLT prediction generating the well-known Kappa distributions \cite{livadiotis2013understanding}.

Thus, understanding the dynamical mechanisms responsible for generating non-Gaussian distributions is not merely a matter of curve-fitting, but conversely it is connected to identifying the nature of the underlying processes, the degree to which the system is driven away from equilibrium, the role of rare but large fluctuations as captured by LDT, and the emergence of collective behavior and intermittency across scales.

A broad spectrum of non-equilibrium systems exhibit similar emerging non-Gaussian statistical features in timeseries \cite{Friedrich2011}. For instance, in space plasma turbulence, non-Gaussianity is one of the most robust and widely documented observational signatures of the energy transfer across scales (i.e., the turbulent cascade). The probability distribution functions (PDFs) of velocity and magnetic field increments are known to become progressively more non-Gaussian as the scale decreases from the energy-injection range toward the dissipative scales, a phenomenon directly linked to the spatial intermittency or non-homogeneity of the structures forming during the turbulent cascade, i.e., the preferential concentration of energy transfer into localized, intense structures such as current sheets, vortex filaments, and so on. This scale-dependent evolution of the PDFs is quantitatively captured by the systematic growth of higher-order statistical moments, as the kurtosis and hyperflatness of increment distributions. Such behavior is intimately connected to the multifractal geometry of the dissipation field. In the magnetohydrodynamic systems relevant to many astrophysical plasmas, additional complexity arises from the interplay between velocity and magnetic field fluctuations, the anisotropy imposed by a background magnetic field, and the presence of compressible effects and kinetic physics at sub-ion scales, all of which leave distinct imprints on the shape of the observed increment PDFs and on the scaling of the structure functions. Moreover, in this framework quantifying and modeling the emergence of non-Gaussian features is essential not only for understanding the statistical mechanics of turbulence, but also for predicting turbulent heating, particle energization, and transport in a wide variety of contexts, where the tails of the distribution, rather than its bulk, often govern the rates of the most physically consequential processes. 

A powerful approach for making progress on this problem is to approach to the scale-to-scale evolution of the fluctuation statistics by means of stochastic process theory. The idea is that an observed coarse-grained dynamics can be modeled as a stochastic process, and a remarkable example of this approach is represented by hydrodynamic turbulence  \cite{friedrich1997description, davoudi1999theoretical, renner2001experimental}, where longitudinal velocity fluctuations at different scales can be described by means of a generalized Langevin equation. When the scale-to-scale dynamics satisfies the Markov condition, the evolution of the PDF is entirely governed by the first two Kramers-Moyal (KM) coefficients, a drift and a diffusion term, in the resulting Fokker-Planck equation (FPE).

An analogous description has been applied to turbulence in space plasmas, and studies confirmed that the FPE is accurate in reproducing the observed PDFs of magnetic field fluctuations in the solar wind \cite{strumik2008testing, benella2022markovian, benella2023linking}. A key empirical result emerging from these analyses is that the magnetic field increments at sub-ion scales are well described by a Langevin process with a linear drift and a quadratic diffusion coefficient, precisely the linear-quadratic (LQ) structure that forms the theoretical core of the present work. For the solar wind, this parameterization of the KM coefficients correctly reproduces the fluctuation statistics at sub-ion scales, where the stationary distribution associated with the LQ Fokker-Planck dynamics is non-Gaussian, heavy-tailed, and potentially asymmetric \cite{benella2022markovian}.

Later Belardinelli et al. (2024) \cite{belardinelli2024derivation} made analytically explicit the emergence of non-Gaussian, heavy-tailed distributions for the magnetic field fluctuations at sub-ion scales by deriving, from the hypothesis of self-similar scaling of magnetic field fluctuations at kinetic scales, the master curve describing their distribution. The result was that PDF tends to be a Kappa distribution whose tail parameter $\kappa$ is connected to the diffusion term. More in detail, the master curve derived in that work can be asymmetric with respect to zero, with the asymmetry controlled by a parameter $\lambda$. This leads to a generalization of the standard Kappa distribution, and to a two-parameter distribution family, $p_{\kappa,\lambda}$, that is the central object of the present paper.

Within the Pearson system, the type IV distribution occupies a distinguished role. It is the unique member supported on the entire real line that simultaneously allows for non-zero skewness and power-law tails with a finite-order moment cutoff. These features are physically relevant when modeling asymmetric, intermittent fluctuations in non-equilibrium dynamics. Despite its theoretical appeal, the Pearson type IV distribution has received comparatively limited attention in the mathematical physics literature \cite{nagahara1999pdf, forman2008pearson, monthus2023large, kar2025call, lanoiselee2025super}, partly due to the complexity of its normalization constant and cumulative distribution function, and partly due to the absence of a systematic derivation rooted in stochastic process theory.

The present paper addresses this gap by providing a self-contained and rigorous derivation of the Pearson type IV distribution as the stationary probability density of a standardized It\^{o} stochastic process whose drift and diffusion coefficient are polynomials of degree one and two in the state variable, respectively, i.e., the LQ case. This polynomial structure can be interpreted as a second-order Taylor expansion of any smooth drift and diffusion around the process mean, making it the natural leading-order approximation for processes whose fluctuations are small relative to their standard deviation. The Pearson type IV distribution thus emerges as the generic stationary distribution of a broad class of weakly non-linear stochastic systems, providing a rigorous theoretical underpinning for the distributions empirically observed in sub-ion solar wind turbulence and other non-equilibrium space plasma environments.

The paper is organized as follows. In Section \ref{Sec:Derivation}, the Fokker-Planck equation for the LQ process is solved exactly under standardization and stationarity constraints, yielding the two-parameter family $p_{\kappa,\lambda}$ in closed form, where $\kappa$ governs the tail behavior and $\lambda$ the asymmetry. Section \ref{Sec:Statistical} derives the main statistical properties of $p_{\kappa,\lambda}$ analytically, including an explicit closed expression for the cumulative distribution function, a recurrence relation for moments of arbitrary order, and explicit formulae for the skewness, kurtosis and mode. Section \ref{Sec:Necessary} establishes necessary and sufficient conditions for the existence of a Pearson type IV stationary solution, and characterizes the full class of drift-diffusion pairs beyond the LQ case that admit $p_{\kappa,\lambda}$ as their stationary distribution. The paper concludes in Section \ref{Sec:Conclusions} with a discussion of the non-standardized case and the interpretation of the shape parameters $\kappa$ and $\lambda$ directly in terms of the diffusion coefficient of the original process.

\section{Derivation}\label{Sec:Derivation}

\begin{figure*}[!ht]
\centering
\includegraphics[width=\textwidth]{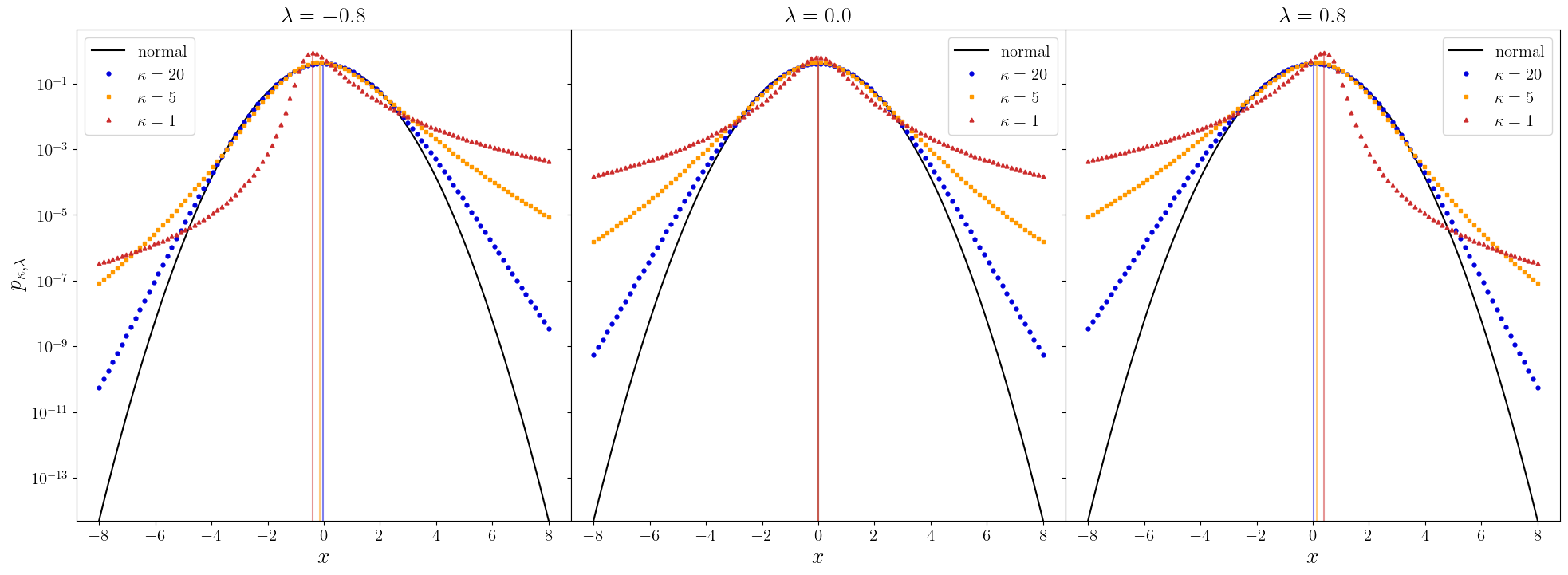}
\caption{Stationary solution to the Fokker-Planck equation for different values of $\kappa$ and $\lambda$. The vertical lines show the mode.}
\label{Fig:distribution}
\end{figure*}
Let $t \geq 0$ denote the independent “time” variable for the Fokker-Planck description of a real Itô stochastic process $(x_t)_{t\geq0}$ satisfying
\begin{equation}
\d{x_t} = m(x_t, t)\d{t} + \sqrt{2d(x_t, t)}\d{W_t}, \label{Eq:langevin}
\end{equation}
where $(W_t)_{t\geq0}$ is a standard Wiener process, with drift $m$ and diffusion coefficient $d$. Let $p_t(x) := \frac{\d{}}{\d{x}}\mathbb{P}(x_t \leq x)$ be the probability density at time $t$ in $x$, satisfying
\begin{equation*}
\int_{-\infty}^{+\infty} p_t(x)\d{x} = 1
\end{equation*}
for any $t \geq 0$. The Fokker-Planck equation corresponding to Eq.~\eqref{Eq:langevin} is then \cite{risken1996fokker}
\begin{equation}
\frac{\partial p_t(x)}{\partial t} = \frac{\partial}{\partial x}\Square{\frac{\partial}{\partial x}(d(x, t)p_t(x)) - m(x, t)p_t(x)}. \label{Eq:FPE}
\end{equation}
By integrating, we immediately get the following condition on the boundary terms for $n = 0$:
\begin{equation}
\Square{x^n\Round{\frac{\partial}{\partial x}(d(x, t)p_t(x)) - m(x, t)p_t(x)}}_{-\infty}^{+\infty} = 0, \label{Eq:boundary_rerms_condition}
\end{equation}
while for $n = 1$ and $n = 2$ it will be assumed to hold as well, for simplicity. We assume the process $(x_t)_{t\geq0}$ to be standardized, that is
\begin{align}
&\int_{-\infty}^{+\infty} x p_t(x)\d{x} = 0, &&\int_{-\infty}^{+\infty} x^2 p_t(x)\d{x} = 1. \label{Eq:standardized}
\end{align}
Then, multiplying Eq.~\eqref{Eq:FPE} by $x^n$ ($n = 1, 2$), integrating and using Eq.~\eqref{Eq:boundary_rerms_condition} we get
\begin{equation}
\begin{split}
&\int_{-\infty}^{+\infty} m(x, t)p_t(x)\d{x} = [d(x, t)p_t(x)]_{-\infty}^{+\infty},\\
&\int_{-\infty}^{+\infty} (d(x, t) + xm(x, t))p_t(x)\d{x} = [xd(x, t)p_t(x)]_{-\infty}^{+\infty}.
\end{split}
\label{Eq:m-d_conditions}
\end{equation}
These are necessary conditions on $m$ and $d$ for the process $(x_t)_{t\geq0}$ to be standardized. In particular, we will eventually focus on the case when the latter are polynomials in $x$ of degree $1$ and $2$, respectively, by writing
\begin{equation}
\begin{split}
m(x, t) &= m_0(t) + m_1(t)x,\\
d(x, t) &= d_0(t) + d_1(t)x + d_2(t)x^2,    
\end{split}
\label{Eq:l-q}
\end{equation}
with $d_2 > 0$ and $4d_0d_2 > d_1^2$. We will refer to this particular choice as the “linear-quadratic” (LQ) case. While a priori we have 5 free time-dependent parameters in the LQ case, standardization (zero mean and unit variance), together with assuming vanishing boundary terms in Eq.~\eqref{Eq:m-d_conditions}, leads to $m_0=0$ and $m_1 = -(d_0 + d_2)$, thus reducing the number of free time-dependent parameters from 5 to 3.

Now, suppose that the process $(x_t)_{t\geq0}$ is stationary. We can then drop time-dependence in $p_t(x) \equiv p(x)$ and Eq.~\eqref{Eq:FPE} gives the condition
\begin{equation*}
\frac{\partial}{\partial x}(d(x, t)p(x)) - m(x, t)p(x) = f(t).
\end{equation*}
Furthermore, $f(t)$ must vanish identically, as a consequence of Eq.~\eqref{Eq:m-d_conditions}. The solution can then be written as
\begin{equation}
p(x) \propto \exp\Round{\int \frac{m(x, t) - \partial d(x, t)/\partial x}{d(x, t)} \d{x}}. \label{Eq:p(x)_general}
\end{equation}
The explicit dependence on $t$ in Eq.~\eqref{Eq:p(x)_general} must be completely reabsorbed, for the process $(x_t)_{t\geq0}$ to be stationary, thus we obtain the condition
\begin{equation}
\frac{m(x, t) - \partial d(x, t)/\partial x}{d(x, t)} \equiv C(x). \label{Eq:master_curve_condition}
\end{equation}
In the special LQ case, that is when $m(x,\cdot)=m_1x=-(d_0+d_2)x$ and $d(x,\cdot)=d_0+d_1x+d_2x^2$, we get $C\equiv C_{\kappa,\lambda}$, with
\begin{equation*}
C_{\kappa,\lambda}(x) := \frac{2[\lambda - (\kappa+1)x]}{2\kappa-1-\lambda^2 + (x-\lambda)^2},
\end{equation*}
where $2\kappa > 1+\lambda^2$, with
\begin{align}
&\kappa := - \frac{m_1(t)}{2d_2(t)} = \frac{1}{2} + \frac{d_0(t)}{2d_2(t)}, &&\lambda := - \frac{d_1(t)}{2d_2(t)}
\label{Eq:kappa-lambda}.
\end{align}
The LQ drift and diffusion coefficients for a stationary standardized process can then be finally parametrized as
\begin{equation}
\begin{split}
m(x, t) &= -2\kappa xd_2(t),\\
d(x, t) &= [2\kappa-1-\lambda^2 + (x-\lambda)^2]d_2(t),
\end{split}
\label{Eq:LQ_kappalambda}
\end{equation}
where $d_2(t)$ remains completely arbitrary (although strictly positive). The parameters $\kappa$ and $\lambda$ respectively regulate the tails and the asymmetry with respect to zero of the resulting stationary distribution:
\begin{equation}
p_{\kappa,\lambda}(x) \propto \frac{\exp\of{- \frac{2\lambda\kappa}{\sqrt{2\kappa - 1 - \lambda^2}}\arctan\of{\frac{x - \lambda}{\sqrt{2\kappa - 1 - \lambda^2}}}}}{\Square{1 + \frac{(x - \lambda)^2}{2\kappa - 1 - \lambda^2}}^{\kappa+1}},
\label{Eq:Kappa-lambda}
\end{equation}
which turns to be a particular form of a Pearson type IV. More precisely,
\begin{equation*}
p_{\kappa,\lambda}(x) := \frac{n_{\kappa,\lambda}}{s_{\kappa,\lambda}} f\of{\kappa, -\frac{\lambda\kappa}{s_{\kappa,\lambda}}; \frac{x - \lambda}{s_{\kappa,\lambda}}},
\end{equation*}
where
\begin{equation}
n_{\kappa,\lambda} := \frac{4^{\kappa}(2\kappa+1)}{\pi}\text{B}\of{\kappa+1 - i\frac{\lambda\kappa}{s_{\kappa,\lambda}}, \kappa+1 + i\frac{\lambda\kappa}{s_{\kappa,\lambda}}} \label{Eq:Kappa-lambda_n}
\end{equation}
is a normalization constant, with $i$ being the imaginary unit and $\text{B}(\alpha, \beta) = \int_0^1 u^{\alpha-1} (1-u)^{\beta-1} \d{u} = \text{B}(\beta, \alpha)$ the well-known beta function, and
\begin{equation}
s_{\kappa,\lambda} := \sqrt{2\kappa - 1 - \lambda^2} \label{Eq:Kappa-lambda_s}
\end{equation}
is a scale factor, while
\begin{equation}
f(\kappa, a; \xi) := \frac{e^{2a\arctan\xi}}{(1 + \xi^2)^{\kappa+1}}. \label{Eq:generating}
\end{equation}
The stationary distribution emerging from the Fokker-Planck dynamics in the LQ case has then the form of a Pearson type IV  \cite{nagahara1999pdf, forman2008pearson, monthus2023large, kar2025call, lanoiselee2025super}. As shown in Fig.~\ref{Fig:distribution}, while a non vanishing $\lambda$ introduces an asymmetry of $p_{\kappa,\lambda}(x)$ with respect to $x = 0$, as quantified by the skewness (see Eq.~\eqref{Eq:skewness} below), the parameter $\kappa$ regulates the power law convergence of the tails and hence the non-existence of the moments above a certain order, as shown in the next section. The condition $2\kappa > 1+\lambda^2$, reflecting the positivity of the diffusion coefficient $d$ in Eq.~\eqref{Eq:LQ_kappalambda}, is actually required for the scale factor $s_{\kappa,\lambda}$ to be real and strictly positive, but can be slightly relaxed in the usage of the “generating” function defined in Eq.~\eqref{Eq:generating}.

\section{Statistical properties}\label{Sec:Statistical}

The identity
\begin{equation}
I(\kappa, a) := \int_{-\infty}^{+\infty} f(\kappa, a; \xi) \d{\xi} = \frac{\pi}{4^{\kappa}(2\kappa+1)\text{B}(\kappa+1+ia, \kappa+1-ia)} \qquad (\kappa > - \tfrac{1}{2})
\label{Eq:integral}
\end{equation}
has been used to define the normalization constant in Eq.~\eqref{Eq:Kappa-lambda_n} as $n_{\kappa,\lambda} = 1/I(\kappa, - i\frac{\lambda\kappa}{s_{\kappa,\lambda}})$. More generally, with the aid of the following result:
\begin{multline*}
c(\kappa, a; \xi) := \int_{-\infty}^{\xi} f(\kappa, a; s) \d{s} \qquad (\kappa > - \tfrac{1}{2})\\
= f(\kappa, a; \xi) \frac{2\operatorname{Im}\Square{{}_2F_1\of{1, \kappa+1+ia; -\kappa+ia; e^{i(\pi-2\arctan\xi)}}} - \xi}{2\kappa+1} + I(\kappa, a)\frac{1 - \operatorname{Re}[\coth(\pi a+i\pi\kappa)]}{2}\\
= 2^{-(2\kappa+1)}\operatorname{Im}\Square{\text{B}_{\frac{1}{2}+\frac{i\xi}{2}}(-\kappa-ia, -\kappa+ia)} + \frac{I(\kappa, a)}{2}\Square{1 + \frac{\sinh(2\pi a)}{\cos(2\pi\kappa) - \cosh(2\pi a)}}
\end{multline*}
with ${}_2F_1(a, b; c; z)$ being the ordinary (Gaussian) hypergeometric function and $\text{B}_z(\alpha, \beta)$ the incomplete beta function, the cumulative distribution function can be computed as
\begin{equation*}
\int_{-\infty}^x p_{\kappa,\lambda}(y) \d{y} = n_{\kappa,\lambda}c\of{\kappa, -\frac{\lambda\kappa}{s_{\kappa,\lambda}}; \frac{x - \lambda}{s_{\kappa,\lambda}}}.
\end{equation*}
By construction, the function $p_{\kappa,\lambda}$, as a distribution of a standardized random variable, satisfies the conditions in Eq.~\eqref{Eq:standardized} for any value of $\kappa$ and $\lambda$ (provided $2\kappa > 1+\lambda^2$). The identities
\begin{equation*}
\frac{1}{I(\kappa, a)} \int_{-\infty}^{+\infty}\xi f(\kappa, a; \xi) \d{\xi} = \frac{a}{\kappa} \qquad (\kappa > 0)
\end{equation*}
and
\begin{equation}
\frac{1}{I(\kappa, a)} \int_{-\infty}^{+\infty} \Round{\xi - \frac{a}{\kappa}}^2 f(\kappa, a; \xi) \d{\xi} = \frac{1+a^2/\kappa^2}{2\kappa - 1} \qquad (\kappa > \tfrac{1}{2}) \label{Eq:integral_2}
\end{equation}
then follow for any real $a$. Furthermore, we can write, for $n \geq 2$ and $2\kappa > n-1$,
\begin{multline*}
\int_{-\infty}^{+\infty} \Round{\xi-\frac{a}{\kappa}}^{n-2} f(\kappa-1, a; \xi)\d{\xi} = \frac{1}{n-1} \int_{-\infty}^{+\infty} \frac{\d{}}{\d{\xi}}\Square{\Round{\xi-\frac{a}{\kappa}}^{n-1}} f(\kappa-1, a; \xi)\d{\xi}\\
= - \frac{1}{n-1} \int_{-\infty}^{+\infty} \Round{\xi-\frac{a}{\kappa}}^{n-1} \frac{\d{f(\kappa-1, a; \xi)}}{\d{\xi}} \d{\xi} = \frac{2\kappa}{n-1} \int_{-\infty}^{+\infty} \Round{\xi-\frac{a}{\kappa}}^n f(\kappa, a; \xi) \d{\xi},
\end{multline*}
where the identity $\d{f(\kappa-1, a; \xi)}/\d{\xi} = 2(a - \kappa\xi)f(\kappa, a; \xi)$ has been used. Consequently, by expanding as
\begin{equation*}
\Round{\xi-\frac{a}{\kappa}}^{n-2} = \sum_{m=0}^{n-2}\frac{(n-2)!}{m!(n-2-m)!}\Round{\xi-\frac{a}{\kappa-1}}^m\Round{\frac{a/\kappa}{\kappa-1}}^{n-2-m}
\end{equation*}
and using that $I(\kappa, a) = \frac{\kappa(\kappa-1/2)}{\kappa^2+a^2}I(\kappa-1, a)$, which is a direct consequence of using the identity $\text{B}(\alpha+1, \beta+1) = \frac{\alpha\beta}{(\alpha+\beta)(\alpha+\beta+1)}\text{B}(\alpha, \beta)$ in Eq.~\eqref{Eq:integral}, we obtain the following useful recurrence relation:
\begin{multline}
\frac{1}{I(\kappa, a)} \int_{-\infty}^{+\infty} \Round{\xi-\frac{a}{\kappa}}^n f(\kappa, a; \xi) \d{\xi} \qquad (n \geq 2, \kappa > \tfrac{n-1}{2})\\
= \frac{1+a^2/\kappa^2}{2\kappa-1} \sum_{m=0}^{n-2}\frac{(n-1)!}{m!(n-2-m)!}\Round{\frac{a/\kappa}{\kappa-1}}^{n-2-m} \frac{1}{I(\kappa-1, a)}\int_{-\infty}^{+\infty} \Round{\xi-\frac{a}{\kappa-1}}^m f(\kappa-1, a; \xi)\d{\xi}. \label{Eq:recurrence_n}
\end{multline}
The result in Eq.~\eqref{Eq:integral_2} is recovered as a particular case. Notice that Eq.~\eqref{Eq:recurrence_n} is only apparently singular at $\kappa = 1$, since in that case it can be applied for $n < 3$ only, which means $n = 2$ and the sum is just the term for $m = 0$. Moments of any order $n \geq 3$ can then be computed using the recurrence relation Eq.~\eqref{Eq:recurrence_n}. For instance, third and fourth order moments respectively follow to be
\begin{equation*}
\frac{1}{I(\kappa, a)} \int_{-\infty}^{+\infty} \Round{\xi-\frac{a}{\kappa}}^3 f(\kappa, a; \xi) \d{\xi} = \frac{2a/\kappa}{\kappa-1}\frac{1+a^2/\kappa^2}{2\kappa-1} \qquad (\kappa > 1)
\end{equation*}
and
\begin{equation*}
\frac{1}{I(\kappa, a)} \int_{-\infty}^{+\infty} \Round{\xi-\frac{a}{\kappa}}^4 f(\kappa, a; \xi) \d{\xi} = \frac{3}{\kappa-1} \frac{1+a^2/\kappa^2}{2\kappa-1} \frac{\kappa-1+(\kappa+3)a^2/\kappa^2}{2\kappa-3} \qquad (\kappa > \tfrac{3}{2}),
\end{equation*}
leading to the skewness
\begin{equation}
\int_{-\infty}^{+\infty} x^3 p_{\kappa,\lambda}(x)\d{x}= - \frac{2\lambda}{\kappa-1} \qquad (\kappa > 1) \label{Eq:skewness}
\end{equation}
and the kurtosis
\begin{equation*}
\int_{-\infty}^{+\infty} x^4 p_{\kappa,\lambda}(x)\d{x} = \frac{3}{\kappa-1}\Round{\kappa + \frac{1+4\lambda^2}{2\kappa-3}} \qquad (\kappa > \tfrac{3}{2}),
\end{equation*}
respectively. The distribution $p_{\kappa,\lambda}$ is unimodal, its mode, that is the value of $x$ such that $p_{\kappa,\lambda}(x)$ is maximum, being given by
\begin{equation*}
\operatornamewithlimits{argmax}_{-\infty<x<+\infty} p_{\kappa,\lambda}(x) = \frac{\lambda}{\kappa + 1}.
\end{equation*}
Furthermore, for large $\kappa$ we obtain the standard normal distribution, regardless of the value of $\lambda$:
\begin{equation*}
\lim_{\kappa\to+\infty} p_{\kappa,\lambda}(x) = \frac{e^{-x^2/2}}{\sqrt{2\pi}}.
\end{equation*}

\section{Necessary and sufficient conditions}\label{Sec:Necessary}

The time-independence of the ratio in Eq.~\eqref{Eq:master_curve_condition} is a necessary condition for the standardized process $(x_t)_{t\geq0}$ to be stationary. Conversely, if the ratio in Eq.~\eqref{Eq:master_curve_condition} is found to be time-independent, then the non-stationary Fokker-Planck equation \eqref{Eq:FPE} becomes
\begin{equation}
\frac{\partial p_t(x)}{\partial t} = \frac{\partial}{\partial x}\Square{d(x, t)\Round{\frac{\partial p_t(x)}{\partial x} - C(x)p_t(x)}}, \label{Eq:non-stationary_FPE_C}
\end{equation}
which admits the stationary solution $p_t(x) \equiv p(x) \propto \exp\of{\int C(x)\d{x}}$. The time-independence condition \eqref{Eq:master_curve_condition} is then sufficient for a stationary solution to exist. Furthermore, in the LQ case such a condition translates into Eq.~\eqref{Eq:kappa-lambda}, which is then necessary for a LQ standardized process $(x_t)_{t\geq0}$ to be stationary and the distribution \eqref{Eq:Kappa-lambda} to emerge. We can wonder if the LQ assumptions in Eq.~\eqref{Eq:l-q} are the only possibility to let the particular Pearson type IV distribution in Eq.~\eqref{Eq:Kappa-lambda} emerge from the Fokker-Planck dynamics. In other words, does the existence of the stationary solution $p_t \equiv p_{\kappa,\lambda}$ imply linear drift and quadratic diffusion? The answer is no. For, assuming $p_t \equiv p_{\kappa,\lambda}$, for some $\kappa$ and $\lambda$, with $2\kappa > 1 + \lambda^2$, we get the relation
\begin{equation*}
m(x, t) = \frac{\partial d(x, t)}{\partial x} + \frac{2[\lambda - (\kappa+1)x]}{s_{\kappa,\lambda}^2 + (x-\lambda)^2}d(x, t),
\end{equation*}
or equivalently
\begin{equation*}
d(x, t) = \frac{1}{p_{\kappa,\lambda}(x)} \int m(x, t)p_{\kappa,\lambda}(x)\d{x}.
\end{equation*}
Thus, in principle, any couple of functions $m$ and $d$ satisfying the previous relations allows the existence of the stationary solution $p_t \equiv p_{\kappa,\lambda}$. For instance, even if a quadratic expression for the diffusion is taken, as in Eq.~\eqref{Eq:l-q}, the resulting drift
\begin{equation*}
m(x, t) = d_1(t) + 2\lambda\frac{d_0(t) + d_1(t)x + d_2(t)x^2}{s_{\kappa,\lambda}^2 + (x-\lambda)^2} + 2x\Square{d_2(t) - (\kappa+1)\frac{d_0(t) + d_1(t)x + d_2(t)x^2}{s_{\kappa,\lambda}^2 + (x-\lambda)^2}}
\end{equation*}
is not necessarily linear, while it reduces to the linear expression of Eq.~\eqref{Eq:l-q} if, in addition, $d_0$, $d_1$ and $d_2$ are such that Eq.~\eqref{Eq:kappa-lambda} hold. Analogously, linear drift does not necessarily give quadratic diffusion. In any case, the shape parameters can be computed by
\begin{equation*}
\begin{split}
\kappa &:= - 1 + \frac{1}{2}\lim_{x\to\infty} x\frac{\partial d(x, t)/\partial x - m(x, t)}{d(x, t)},\\
\lambda &:= \Round{\kappa - \frac{1}{2}}\lim_{x\to0}\frac{m(x, t) - \partial d(x, t)/\partial x}{d(x, t)},
\end{split}
\end{equation*}
which can be taken as the general definitions of $\kappa$ and $\lambda$ every time the particular Pearson type IV distribution $p_{\kappa,\lambda}$ is observed for the stationary process defined by Eq.~\eqref{Eq:langevin}.

\section{Conclusions}\label{Sec:Conclusions}

The particular Pearson type IV distribution $p_{\kappa,\lambda}$ in Eq.~\eqref{Eq:Kappa-lambda} is the stationary probability distribution of a standardized Itô stochastic process $(x_t)_{t\geq0}$ evolving according to Eq.~\eqref{Eq:langevin} for a large class of drift $m$ and diffusion coefficient $d$. This includes the LQ case \eqref{Eq:l-q}, which can be seen as a Taylor expansion of any drift and diffusion coefficient around the average value $x = 0$. With this interpretation, the definitions in Eq.~\eqref{Eq:kappa-lambda} can be straightforwardly rewritten as
\begin{equation}
\begin{split}
\kappa &:= - \lim_{x\to0} \frac{\partial m(x, t)/\partial x}{\partial^2d(x, t)/\partial x^2} = \frac{1}{2} + \lim_{x\to0} \frac{d(x, t)}{\partial^2d(x, t)/\partial x^2},\\
\lambda &:= - \lim_{x\to0} \frac{\partial d(x, t)/\partial x}{\partial^2d(x, t)/\partial x^2}.
\end{split}
\label{eq:stand-kappa-lambda}
\end{equation}
A non-standardized process $(X_t)_{t\geq0}$, for which
\begin{equation}
\d{X_t} = M(X_t, t)\d{t} + \sqrt{2D(X_t, t)}\d{W_t} \label{Eq:Langevin}
\end{equation}
for some drift $M$ and the diffusion coefficient $D$, can always be standardized as
\begin{equation*}
x_t := \frac{X_t - \mu_t}{\sigma_t},
\end{equation*}
where $\mu_t := \mathbb{E}(X_t)$ and $\sigma_t := \sqrt{\mathbb{E}(X_t^2) - \mu_t^2}$. In this case \cite{belardinelli2024derivation},
\begin{equation*}
\begin{split}
m(x, t) &:= \frac{M(\mu_t+\sigma_tx, t) - \frac{\d{\mu_t}}{\d{t}} - \frac{\d{\sigma_t}}{\d{t}}x}{\sigma_t},\\
d(x, t) &:= \frac{D(\mu_t+\sigma_tx, t)}{\sigma_t^2}.
\end{split}
\end{equation*}
It follows that the standardized process $(x_t)_{t\geq0}$ is LQ every time $(X_t)_{t\geq0}$, and vice versa. Consequently, any stochastic process $(X_t)_{t\geq0}$ evolving according to Eq.~\eqref{Eq:Langevin}, whose oscillations around its ensemble average $\mu_t$ are small with respect its standard deviation $\sigma_t$ and whose standardized version is stationary, is statistically described by the following Pearson type IV time-dependent probability density:
\begin{equation*}
P_{\kappa,\lambda}(X, t) := \frac{n_{\kappa,\lambda}}{s_{\kappa,\lambda}\sigma_t} \frac{\exp\of{- \frac{2\lambda\kappa}{s_{\kappa,\lambda}}\arctan\of{\frac{X - \mu_t - \lambda\sigma_t}{s_{\kappa,\lambda}\sigma_t}}}}{\Square{1 + \Round{\frac{X - \mu_t - \lambda\sigma_t}{s_{\kappa,\lambda}\sigma_t}}^2}^{\kappa+1}},
\end{equation*}
with $n_{\kappa,\lambda}$ and $s_{\kappa,\lambda}$ respectively given in Eq.~\eqref{Eq:Kappa-lambda_n} and Eq.~\eqref{Eq:Kappa-lambda_s}, while
\begin{equation*}
\begin{split}
\kappa &= \frac{1}{2} + \frac{1}{\sigma_t^2}\lim_{X\to\mu_t} \frac{D(X, t)}{\partial^2D(X, t)/\partial X^2},\\
\lambda &= - \frac{1}{\sigma_t}\lim_{X\to\mu_t} \frac{\partial D(X, t)/\partial X}{\partial^2D(X, t)/\partial X^2}.
\end{split}
\end{equation*}

From the physical perspective, a key insight emerging from our results is that the tails of the stationary PDF are regulated by the competition between the average deterministic drift, representing the global restoring force relaxing the system towards its mean state, and the state-dependent fluctuations, representing the local stochastic feedback. This is evident by inspecting Eqs. \eqref{eq:stand-kappa-lambda}, where the shape parameter $\kappa$ expresses the ratio between the linear term of the drift in the Taylor expansion, i.e., $\partial m/\partial x|_{x=0}$, and the non-linear diffusion term, i.e., $\partial^2 d/\partial x^2|_{x=0}$. In the limit of constant diffusion coefficient, the shape parameter $\kappa\to\infty$ and the stationary PDF converges to the Gaussian. Heavy tails are thus the signatures of multiplicative noise \cite{vo2026kinetic}, leading to intermittent extreme events often observed in non-equilibrium timeseries. Conversely, the asymmetry parameter $\lambda$ is only determined by diffusion parameters, being the ratio between local gradient and curvature of $d$. Since the standardized drift is strictly linear, it corresponds to a harmonic potential $V(x)\propto x^2$ and the asymmetry is a merely noise-induced phenomenon. A nonzero diffusion local gradient $\partial d/\partial x|_{x=0}$ introduces a linear term in the FPE that can be viewed as a spurious drift, breaking the symmetry of the state-dependent fluctuations and leading to skewed PDFs without requiring any asymmetry of the deterministic restoring force.

To summarize, the statistics of large fluctuations in non-equilibrium stationary states is regulated by the interplay between state-dependent drift and diffusion functions, which thus embed any departure from Gaussianity, thus providing a generalized framework to model asymmetric heavy-tailed phenomena across diverse non-equilibrium systems. 
The Pearson type IV PDF provides a theoretical benchmark whenever the local state-dependent diffusivity can be leading-order approximated by a second-degree polynomial around the mean. This is applicable across a wide spectrum of non-equilibrium systems, ranging from driven colloidal suspensions to hydrodynamic and space plasma turbulence, active matter, and financial time series.



\bibliographystyle{unsrt}
\bibliography{Belardinelli.bib}

\end{document}